\documentclass[dvipsnames, twocolumn]{aastex701}

\newcommand{\ee}[1]{\mbox{${} \times 10^{#1}$}}
\newcommand{\eten}[1]{\mbox{$10^{#1}$}}
\newcommand{\as}{\mbox{\arcsec}}
\newcommand{\degree}{\mbox{$^{\circ}$}}

\newcommand{\water}{H$_2$O}

\newcommand{\cooo}{C$^{18}$O}

\newcommand{\hh}{\mbox{{\rm H}$_2$}}
\newcommand{\nn}{\mbox{{\rm N}$_2$}}

\newcommand{\nnhp}{$\mathrm{N_{2}H^+}$}

\newcommand{\hcop}{\mbox{\rm HCO}$^+$}
\newcommand{\hthcop}{$\mathrm{H^{13}CO^+}$}

\newcommand{\hhhp}{\mbox{\rm H}$_3^+$}
\newcommand{\hhdp}{\mbox{\rm H}$_2$D$^+$}
\usepackage{amsmath}

\shorttitle{Ionization in Protostellar Envelopes}
\shortauthors{Schwarz et al.}
\graphicspath{{./}{figures/}}

\begin{document}

\title{Evidence of Enhanced Ionization in Protostellar Envelopes}

\author[0000-0002-6429-9457]{K. R. Schwarz}
\affiliation{Max-Planck-Institut für Astronomie, Königstuhl 17, 69117 Heidelberg, Germany}
\affiliation{Steward Observatory, University of Arizona, 933 North Cherry Avenue, Tucson, AZ 85721, USA}
\email[show]{kschwarz@arizona.edu} 

\author[0000-0003-1104-4554]{S. Maret}
\affiliation{Univ. Grenoble Alpes, CNRS, IPAG, 38000 Grenoble, France}
\email{sebastien.maret@univ-grenoble-alpes.fr}

\author[0000-0002-8120-1765]{M. R. A. Wells}
\affiliation{Max-Planck-Institut für Astronomie, Königstuhl 17, 69117 Heidelberg, Germany}
\email{wells@mpia.de}

\author[0000-0002-8120-1765]{C. Gieser}
\affiliation{Max-Planck-Institut für Astronomie, Königstuhl 17, 69117 Heidelberg, Germany}
\email{gieser@mpia.de}

\author[0000-0003-0046-6217]{A. Belloche}
\affiliation{Max-Planck-Institut für Radioastronomie, Auf dem Hügel 69, 53121 Bonn, Germany}
\email{belloche@mpifr-bonn.mpg.de}

\author[0000-0002-3413-2293]{P. Andre}
\affiliation{Laboratoire d’Astrophysique (AIM) Université Paris-Saclay, Université Paris Cité, CEA, CNRS, AIM 91191 Gif-sur-Yvette, France;}
\email{pandre@cea.fr}

\author[0000-0003-1514-3074]{C. Codella}
\affiliation{INAF, Osservatorio Astrofisico di Arcetri, Largo E. Fermi 5, 50125 Firenze, Italy}
\affiliation{Univ. Grenoble Alpes, CNRS, IPAG, 38000 Grenoble, France}
\email{claudio.codella@inaf.it}

\begin{abstract}
Ionization is a major driver of both physical and chemical evolution in protostellar systems. Recent observations reveal substantial chemical processing in protoplanetary disks by the time the surrounding envelope has cleared. Thus, physical conditions during the preceeding phase, when an infalling envelope of material is still present, are crucial for determining the extent of chemical processing at early stages. 
We used observations of \hthcop\ and \cooo\ from the Northern Extended Millimeter Array (NOEMA) and IRAM 30m telescope to constrain the ionization rate in the envelopes of three Class 0 protostars: NGC-1333 IRAS4A, L1448-C, and L1157. We find ionization rates in the range $\zeta = \eten{-16}-\eten{-13}$ s$^{-1}$, several orders of magnitude above the ionization rate of $\zeta =  6\ee{-17}$ s$^{-1}$ in the diffuse interstellar medium. This supports the idea that ionization driven chemistry is more efficient at earlier stages ($< 10^5$ years) of protostellar evolution.
\end{abstract}

\section{Introduction}
During star and planet formation the level of ionization has important implications for both dynamical and chemical processes. 
The level of ionization determines how well gas is coupled to the magnetic fields, and thus determines the importance of magnetohydrodynamics (MHD). In turn, MHD affects processes such as disk formation, mass transport through the disk, and accretion onto the central protostar \citep{Tsukamoto23,Lesur23}.

Young stellar objects (YSOs) are divided into classes based on their evolutionary stage \citep{Adams87,Andre95,Furlan16}.
In Class 0 protostellar systems the envelope mass, $M_{env}$, is higher than the mass of the stellar embryo, $M_{*}$, while in Class I systems $M_{env} < M_{*}$ and Class II systems have no evidence for an envelope in their SED \citep{Andre00}.
In both disks and envelopes, much of the gas phase chemistry is driven by ion-neutral reactions \citep{Aikawa99,Jorgensen04}. The abundance of ions depends on the degree of ionization, which in turn depends on the flux of energetic particles and high energy photons.

Surveys with the Atacama Large Millimeter/submillimeter Array reveal that many Class II sources have a low CO abundance relative to dust \citep{Ansdell16,Eisner16,Long17,Barenfeld17,Trapman25}.
In some disks, this ratio is over 100 times lower than that observed in the dense ISM \citep{Schwarz16,Anderson19,Zhang20}. This low ratio is present in disks as young as one million years \citep{Zhang21}. 
The abundances of other volatile species, such as warm water vapor and atomic carbon, are sometimes also lower than expected \citep{Kama16,Du17,Notsu19,Pascucci23,Facchini24,Nakasone26}.

This suggests that volatile molecules have been chemically reprocessed into less volatile species and/or trapped within growing dust grains.
Chemical models of Class II T Tauri disks show that chemistry can significantly lower the gas phase carbon abundance in a few million years, but only with a cosmic ray ionization rate of 1.3\ee{-17} s$^{-1}$ \citep{Reboussin15,Eistrup16,Bosman18,Schwarz18}. This rate is based on spacecraft measurements at a distance of 60 au from the Sun \citep{Webber98}.

Observations of molecular ions in several Class II disks suggest lower rates of \eten{-19}-\eten{-18} s$^{-1}$ \citep{Cleeves15,Seifert21,Aikawa21,Long24}.
Low ionization rates in Class II disks are thought to be due to winds from the central star modulating the incoming Galactic cosmic rays \citep{Cleeves13}. Since stellar winds should be ubiquitous at this stage, the cosmic ray ionization rate in all Class II disks could be too low to facilitate rapid chemical processing.
Rotation induced shearing of the disk magnetic field can also increase the path-length of cosmic rays traveling through the disk, resulting in a lower cosmic ray flux near the midplane \citep{Fujii22}.

Observations of several disks still embedded in their natal envelope, e.g., Class 0/I sources, reveal disk CO gas abundances similar to that in the ISM, while significant depletion is common by one million years \citep{vantHoff18,Booth20,Zhang20,RuizRodriguez25}. The envelopes of these systems also show evidence of reduced CO abundance \citep{Anderl16}. Meanwhile, cold \water\ vapor can already be depleted by the Class I stage \citep{Harsono20,Facchini24}. Taken together, these studies suggest that volatiles are removed from the gas quickly, when the age of the protostellar system is less than one million years.

The ionization environment for younger, embedded systems is likely more conducive to chemical reprocessing.
Recent studies of ionization in diffuse clouds find an average ionization rate of $\zeta\approx6\times10^{-17}$ s$^\mathrm{-1}$, roughly nine times lower than previous estimates \citep{Obolentseva24}. The ionization rate further decreases in dense starless cores ($1.3\times 10^{-18}$ to $8.5\times 10^{-17}$ s$^\mathrm{-1}$), as Galactic cosmic rays loose energy \citep{Sabatini20,Sabatini23,Redaelli25}. However, once protostars are present the local ionization rate appears to increase by several orders of magnitude \citep{Pineda24}.

Models show that shocks along protostellar jets and at the protostellar surface can accelerate cosmic rays, increasing the local ionization rate \citep{Padovani16,Gaches18}.
Observations of two Class 0 envelopes imply a cosmic ray ionization rate of order \eten{-14} s$^{-1}$ \citep{Ceccarelli14, Favre17, Cabedo23, Luo24}. The analysis of molecular ions in the protostellar bow-shock L1157-B1 results in an ionization rate of 3\ee{-16} s$^{-1}$ \citep{Podio14, Luo24}. Additionally, the detection of non-thermal emission at the jet base of some protostellar sources suggests cosmic ray acceleration \citep{Tychoniec18,Sanna19,Bouvier21,Boyden24}. Conversley, chemical modeling of the Class 0/I source L1527 IRS supports a reduced rate of \eten{-18} s$^{-1}$\citep{vantHoff22}. 
If high ionization rates are found to be common in young systems, it will have major implications for protostellar envelope chemistry as well as the initial chemical abundances in protoplanetary disks.

The CALYPSO\footnote{\url{http://irfu.cea.fr/Projets/Calypso/About_CALYPSO.html}} survey of Class 0 sources was undertaken at the IRAM Plateau de Bure Interferometer, yielding a number of interesting results constraining the physics and chemistry in Class 0 protostellar systems at high resolution \citep[e.g.,][]{Maret14,Santangelo15,Maury19,Belloche20}.
As part of this survey, \citet{Anderl16} explored the question of the chemical abundances and relationship between CO and \nnhp\ in the envelopes of four protostellar systems. 
Using dust as a tracer of the gas mass, they determined that on small scales ($<3.4''$, 800 au), where the temperature is greater than 20~K and CO should be in gas, some envelope CO appears to be missing; a result similar to the more evolved Class II disk systems. 
In this work we used archival CALYPSO observations of \cooo\ as well as additional observations of \hthcop\ and \cooo\ from the Northern Extended Millimeter Array (NOEMA) and the IRAM 30m telescope to constrain the ionization rate in three protostellar envelopes previously studied by \citet{Anderl16}. The basic properties of the observed sources are given in Table~\ref{tab:sourceprops}.

\section{Observations \& Data Reduction}
\subsection{NOEMA Observations}
New observations of the \hthcop\ J = 1-0 line were carried out with the IRAM NOEMA Interferometer between October 2016 and April 2018.
NGC 1333-IRAS4A was observed in October 2016 with 8 antennas with baselines in the range 24-240 meters for a total of 9 hours on source. 
We used the Band 1 receivers with a tuning frequency of 86.75 GHz, and the narrow-band correlator with a spectral resolution of 39 kHz and total bandwidth of 20 MHz. In addition to \hthcop, HN$^{\mathrm{13}}$C, HC$^{\mathrm{15}}$N, and SiO were also targeted in this spectral setup. 
Bandpass calibration was performed on 3C84 and 3C454.3. The phase calibrators where 0333+321 and 0234+285.
Calibration and imaging were carried out using the \texttt{CLIC} and \texttt{MAPPING} packages from the \texttt{GILDAS}\footnote{\url{https://www.iram.fr/IRAMFR/GILDAS}} software suite. 
The 3mm continuum, determined using the line free channels in the WideX windows, was subtracted in $uv$ space before imaging the line emission. 

\begin{deluxetable}{lcccc}
\tablecaption{Source Properties}
\tablewidth{1.0\columnwidth} 
\label{tab:sourceprops}
\tablehead{
\colhead{Source} & \colhead{RA} & \colhead{Dec} & \colhead{v$_\mathrm{lsr}^a$} & \colhead{dist.$^b$} \\
 & \colhead{(HMS)} & \colhead{(DMS)} & \colhead{(km s$^{-1}$)}  &\colhead{(pc)}}
\startdata 
IRAS4A & 03:29:10.43 & 31:13:32.20 & 7.2 & 235\\ 
L1448-C & 03:25:38.88 & 30:44:05.30 & 5.2 & 235\\ 
L1157 & 20:39:06.27 & 68:02:15.70  & 2.8 & 325
\enddata 
\tablecomments{
$^a$v$_\mathrm{lsr}$ from \cooo\ 2-1 as derived by \citet{Anderl16}. $^b$Distances taken from \citet{Kristensen12}.
}
\end{deluxetable}

L1448-C and L1157 where observed in April 2018 with 9 antennas in C configuration with baselines in the range 24-704 meters for a total of 2.2 hours and 8 hours on source respectively.  
We used the Band 1 receivers with a tuning frequency of 86.782 GHz, and the new POLYFIX correlator with a spectral resolution of 65 kHz, and bandwidth of 192 MHz.
For the observations of L1448-C, bandpass calibration was performed on 3C454.3, phase calibrators were 0333+321 and J0329+351, and flux calibrators were MWC 349 and 1928+738. For L1157 the bandpass calibrator was 1055+018, the phase calibrators were 2010+723 and 1928+738, and the flux calibrator was MWC 349.
Calibration procedures mirror those carried out for IRAS4A. 

\begin{deluxetable*}{lcccccc}
\tablecaption{Parameters for final datacubes}
\label{tab:dataprops}
\tablehead{
\colhead{Source} & \colhead{Line} & \colhead{Beam} & \colhead{Peak flux} & \colhead{rms} & \colhead{Channel spacing} & \colhead{Velocity range} \\  
 & & \colhead{($\arcsec\times\arcsec$, $\degree$)} & \colhead{(mJy beam$^{-1}$)} & \colhead{(mJy beam$^{-1}$)} & \colhead{(km s$^{-1}$)} & \colhead{(km s$^{-1}$)} 
}
\startdata 
IRAS4A & \hthcop\ J=1-0 &($2.70\times2.35, 197.4$) & 148.5  & 4.5 & 0.135 & 5.99-8.42 \\ 
       & \cooo\ J=2-1 &($0.80\times0.66, 26.7$) & 364.4 & 25.9 & 0.10 & 4.02-8.88 \\ 
L1448-C & \hthcop\ J=1-0 &($2.12\times1.84, -160.5$) & 49.6 & 2.0 & 0.4 & 4.3-6.3\\ 
       & \cooo\ J=2-1 &($0.63\times0.39, 31.2$) & 319.2 & 3.8 & 0.10 & 3.90-7.10\\ 
L1157 & \hthcop\ J=1-0 &($3.31\times2.36, 56.6$) & 108.2 & 3.0 & 0.4 & 0.4-4.0\\ 
       & \cooo\ J=2-1 &($0.62\times0.47, -165.9$) & 288.2 & 2.9 & 0.10 & 0.99-3.93\\ 
\enddata 
\end{deluxetable*}

\begin{figure*}
    \centering
    \includegraphics[width=2\columnwidth]{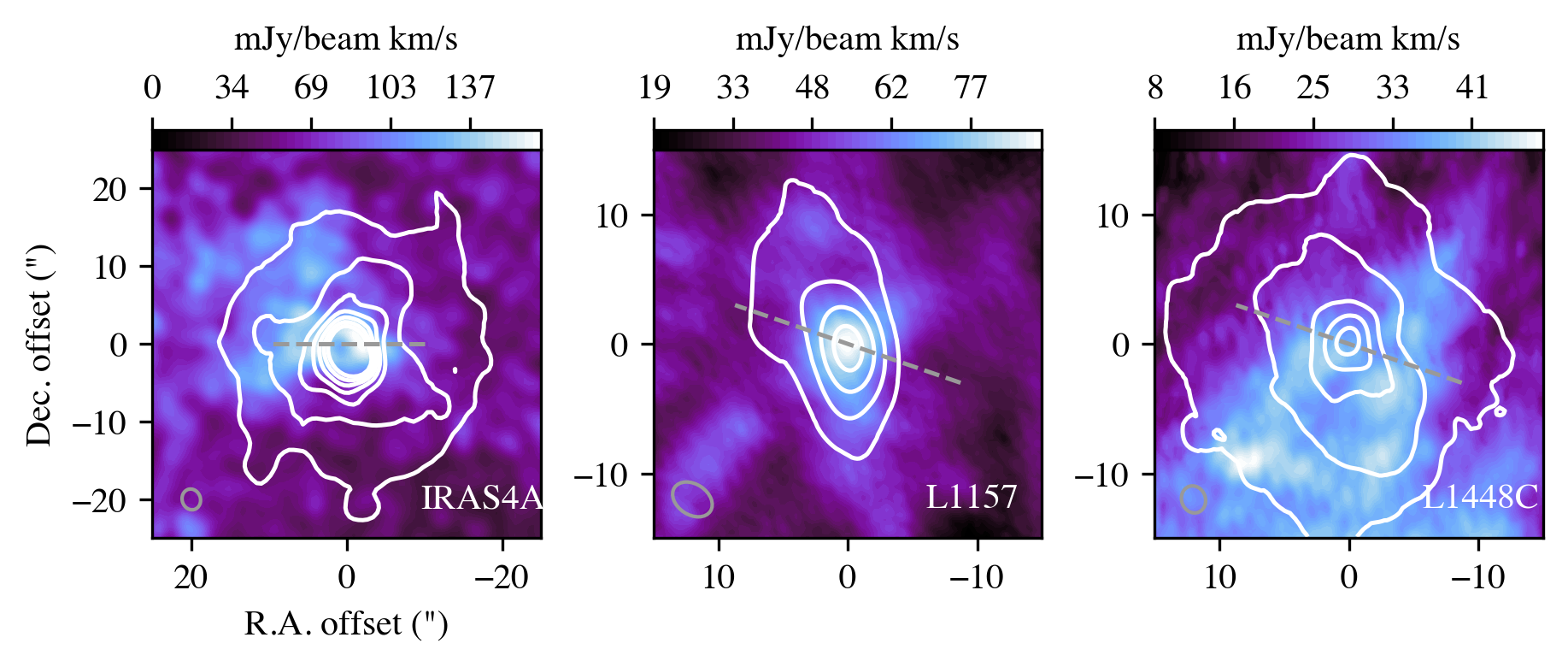}
    \caption{Combined NOEMA+IRAM 30m observations of \hthcop\ 1-0 (background), and \cooo\ 2-1 (contours, convolved to the resolution of the \hthcop) envelope emission toward three protostars: NGC 1333-IRAS4A (left), L1157 (center), and L1448-C (right). Contours start at $3\sigma$ and have $6\sigma$ spacing for IRAS4A and $3\sigma$ spacing for the other sources. The grey dashed lines are perpendicular to the small scale outflows. The grey ellipse in the bottom left corner of each panel show the synthesized beam of the \hthcop\ 1-0.}
    \label{mom0}
\end{figure*}

Additionally, we use archival \cooo\ J = 2-1 observations from the CALYPSO survey. A detailed description of the observations and data reduction process can be found in \citet{Maret20} and \citet{Gaudel20}.

\subsection{IRAM  Observations}
Given the large extent of line emission in protostellar envelopes, zero-spacing observations must be combined with the interferometric NOEMA observations to ensure no flux is lost \citep{Gaudel20,Gieser25}. Observations were carried out with the IRAM 30m telescope using the Eight MIxer Receiver (EMIR) targeting the \hthcop\ 1-0 line at 86.75 GHz with a spectrometer resolution of 0.12 km s$^{-1}$. IRAS4A and L1157 were observed for a total of 8 hours on source while L1448-C was observed for 4 hours. The IRAM 30m telescope observations were carried out in on-the-fly (OTF) mode in July of 2022. The OTF maps have a size of $1.67'\times1.67'$ and were observed in zigzag scanning mode in right ascension and declination. 

The reduction of the single-dish data was carried out using the \texttt{CLASS} package of \texttt{GILDAS}. Baseline subtraction for individual spectra was carried out by fitting a straight line around the \hthcop\ emission line. The antenna temperature ($T^*_A$) was converted to a main beam temperature ($T_{MB}$) using the formula $T_{MB} = \frac{F_{eff}}{B_{eff}}\times T^*_A$, where the forward efficiency $F_{eff} = 0.95$ and the main beam efficiency $B_{eff} = 0.81$.\footnote{\url{https://publicwiki.iram.es/Iram30m telescopeEfficiencies}} The single-dish spectra were then re-binned to match the spectral resolution of the NOEMA data. 

\subsection{Merging data}
\begin{figure}
    \includegraphics[width=\columnwidth]{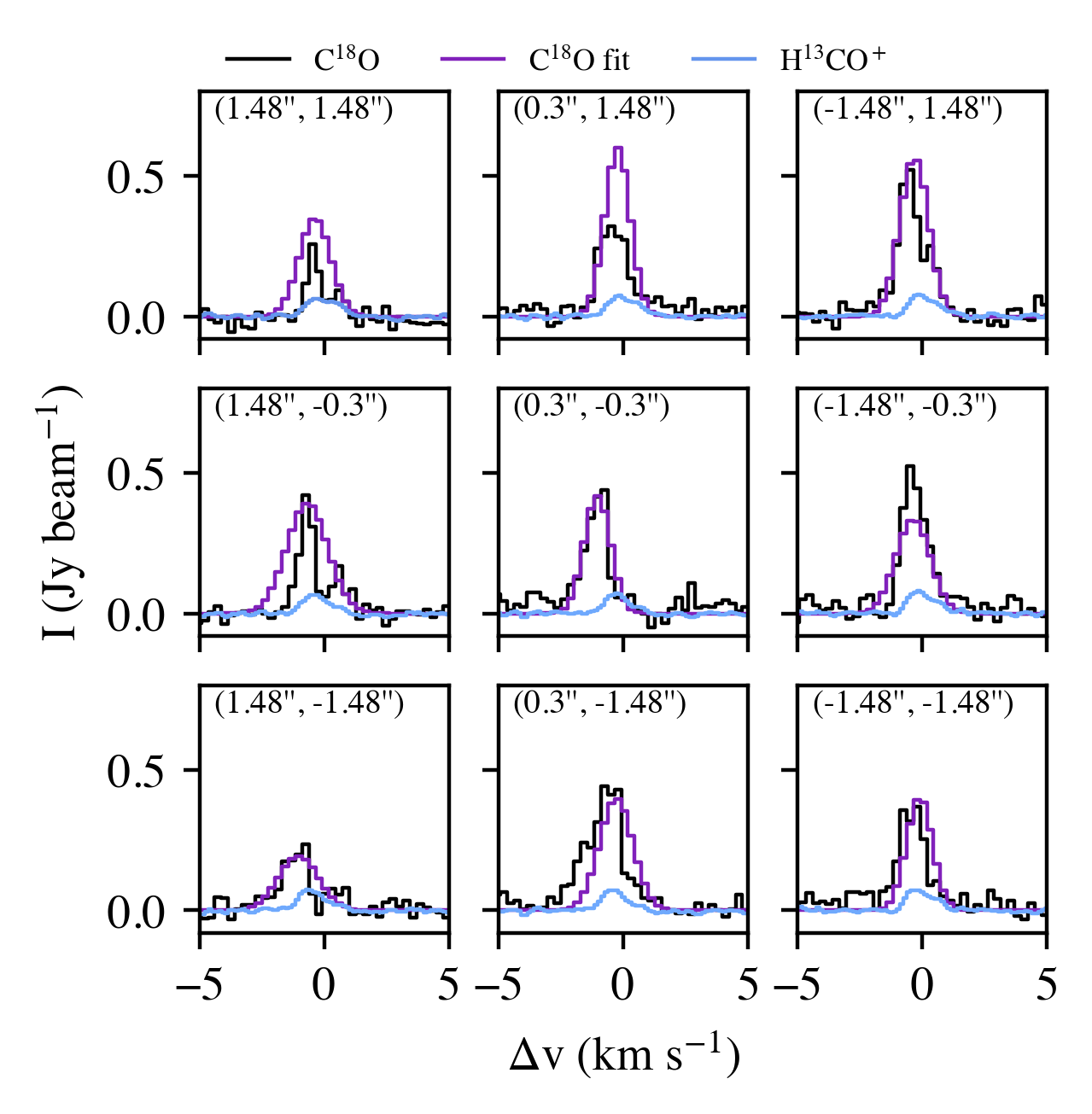}
    \caption{\hthcop\ 1-0 (blue) and \cooo\ 2-1 (black, convolved to the spatial resolution of \hthcop) spectra toward NGC 1333-IRAS4A. The purple line shows the best Gaussian fit to the \cooo\ line correcting for absorption. The spectra in each panel are offset by half $\theta_{maj}$ for the \hthcop\ observations.}
    \label{iras4aspec}
\end{figure}

\begin{figure}
    \includegraphics[width=\columnwidth]{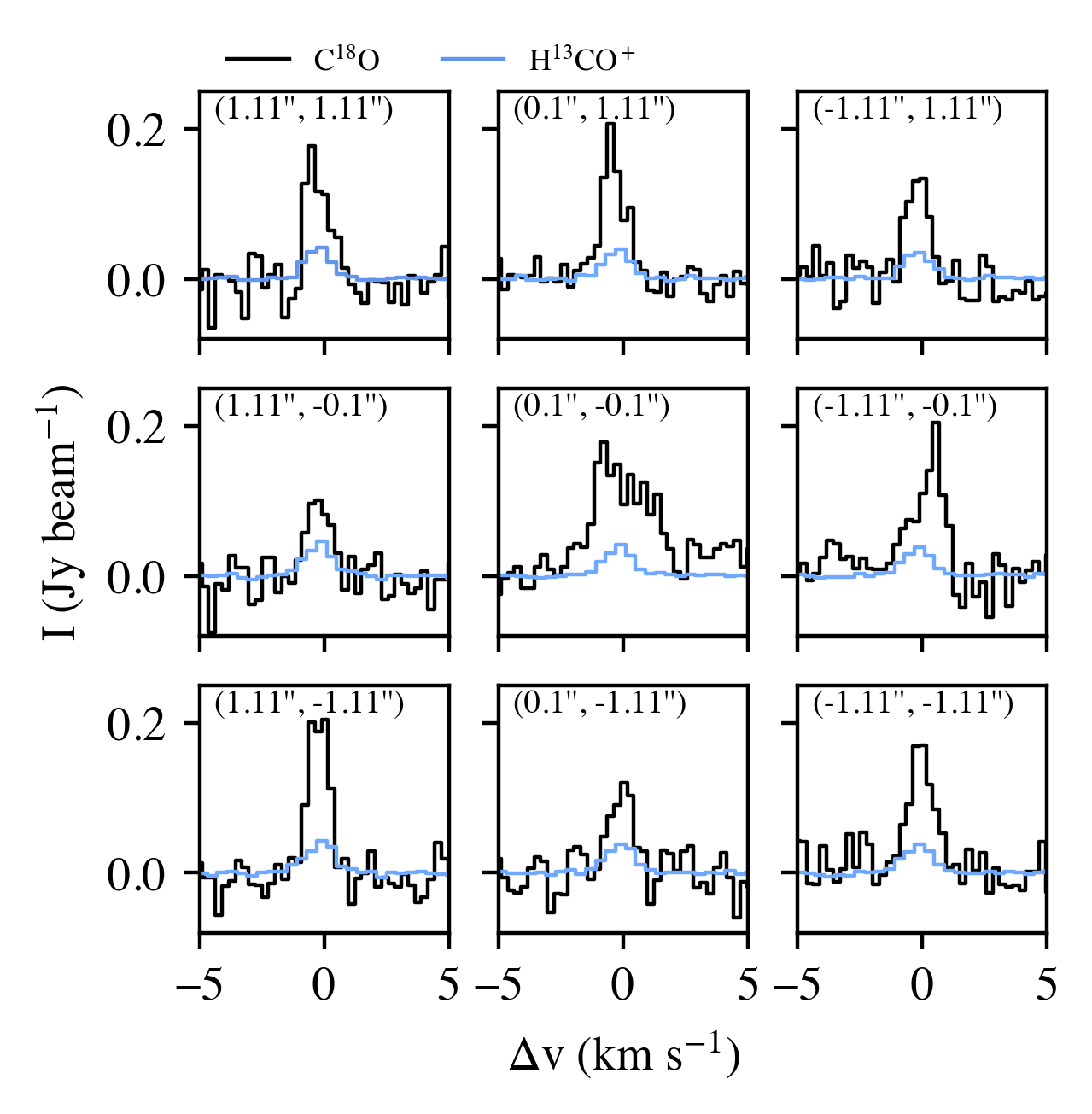}
    \caption{\hthcop\ 1-0 (blue) and \cooo\ 2-1 (black, convolved to the spatial resolution of \hthcop) spectra toward L1448-C. The spectra in each panel are offset by half $\theta_{maj}$ for the \hthcop\ observations.}
    \label{l1448spec}
\end{figure}

\begin{figure}
    \includegraphics[width=\columnwidth]{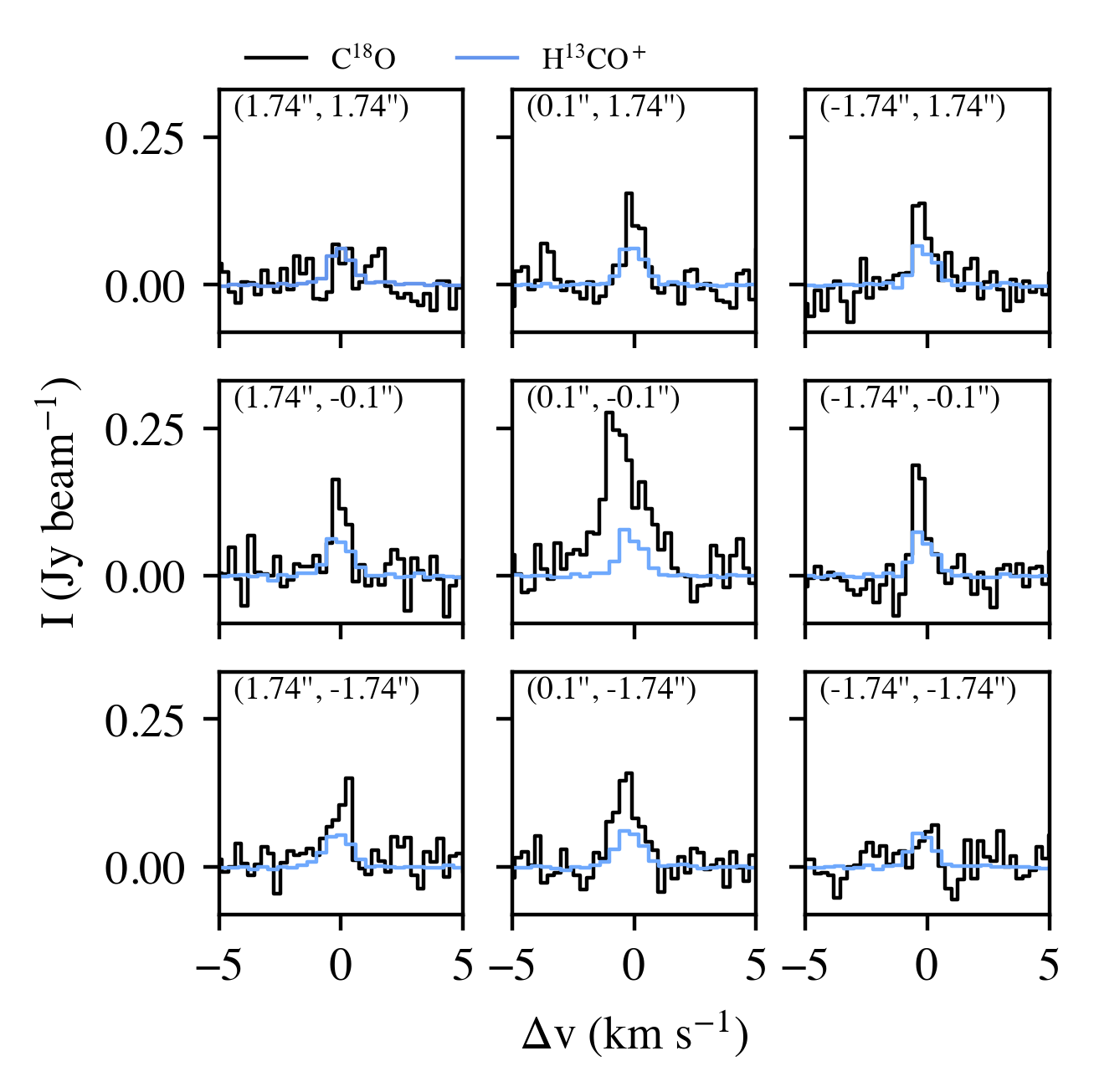}
    \caption{\hthcop\ 1-0 (blue) and \cooo\ 2-1 (black, convolved to the spatial resolution of \hthcop) spectra toward L1157. The spectra in each panel are offset by half $\theta_{maj}$ for the \hthcop\ observations.}
    \label{l1157spec}
\end{figure}

The single-dish 30m telescope and interferometric NOEMA spectral line data were combined using the  \texttt{uv\_short} task in \texttt{MAPPING}. 
For \cooo\ single-dish observations were obtained and combined with the NOEMA data by \citet{Gaudel20}.
For the \hthcop\  observations, the interferometric data were first merged with the single-dish weighting factor set to 1, then imaged using natural weighting, and finally cleaned down to the theoretical noise limit or to a maximum of 5000 clean components per channel using the Hogbom algorithm. The properties of the final merged datacubes for both the new \hthcop\ observations and the archival \cooo, as well as the velocity ranges used to generate integrated intensity maps, are given in Table~\ref{tab:dataprops}. Integrated intensity maps of the \hthcop\ J = 1-0 and \cooo\ J = 2-1 emission are shown in Figure~\ref{mom0} while a selection of spectra for independent pointings are shown in Figures~\ref{iras4aspec}-\ref{l1157spec}.

\subsection{Absorption Correction}
The combined \cooo\ spectra toward IRAS4A follow an inverse P Cygni profile indicative of infall, a feature seen in multiple previous studies of this and other protostellar objects \citep{DiFrancesco01,Belloche06,Kristensen12,vanDishoeck21}.  This must be corrected for in order to derive a \cooo\ column density.  We fit a Gaussian profile to the spectrum in each pixel using the Markov Chain Monte Carlo package \texttt{emcee} \citep{emcee}. The central 0.6 km s$^{-1}$ around the source velocity, where the absorption feature is seen, are not included in the fit. The best fit for the central spectra is shown in Figure~\ref{iras4aspec}. Neither the \hthcop\ in this source, nor the spectra of the other sources, show evidence for self absorption in their line profiles. However, the lower spectral resolution of the \hthcop\ observations toward L1448-C and L1157 could be masking such features. If this is the case, then the intensities used in the below analysis should be considered lower limits.  

\section{Analysis}\label{sec:ana}
Molecular ions, particularly \hcop\ isotopologues, have long been used to derive the ionization fraction in the dense interstellar medium \citep{Caselli98}. Recently, \citet{Redaelli24} demonstrated that the final derivation for measuring ionization from \citet{Caselli98} does not provide accurate measurements in prestellar cores. This is partially due to high levels of CO freeze-out in these objects and partially due to updated measurements of key reaction rates. For the Class 0 protostellar envelopes which are the focus of the current study, CO freeze-out is less of an issue due to the warmer temperatures.  Below, we re-derive the equation for measuring ionization based on \hcop\ and CO observations, following the conventions of \citet{Redaelli24} and using updated reaction rates from the KIDA network \citep{Wakelam24}\footnote{\url{https://kida.astrochem-tools.org/}}.

\hcop\ is formed via a reaction between CO and \hhhp, an ion formed from \hh. The abundance ratio of CO and \hcop\ can be related to the ionization rate assuming steady state chemistry. In the following derivation we assume \hhhp\ reacts primarily with CO and neglect other reactions. 
Chemical models shows this to be a reasonable assumption for gas where both CO and \hcop\ are abundant while \water\ remains largely in the ice \citep{Bergin97,Rodgers03}. Reactions with \nn, which has a signifcantly lower abundance than CO, are ignored \citep{Womack92,Visser11,Furuya18}. 
Setting the formation and destruction rates for \hcop\ equal gives 
\begin{equation}\label{eq:hcop}
\frac{N(\mathrm{HCO^+})}{N(\mathrm{CO})} = \frac{n(\mathrm{H_3^+}) k_{\mathrm{H_3^+}} }{n(\mathrm{e}) \beta_{\mathrm{HCO^+}}}.
\end{equation}
Similarly setting the formation and destruction rates for \hhhp\ equal gives
\begin{equation}\label{eq:h3p}
n(\mathrm{H_3^+})  = \frac{\zeta n(\mathrm{H_2}) }{k_{\mathrm{H_3^+}} n(\mathrm{CO}) + n(\mathrm{e}) \beta_{\mathrm{H_3^+}}}.
\end{equation}
Plugging Equation~\ref{eq:h3p} into Equation~\ref{eq:hcop} gives
\begin{equation}\label{eq:final}
\frac{N(\mathrm{HCO^+})}{N(\mathrm{CO})} = \frac{k_{\mathrm{H_3^+}} \zeta n(\mathrm{H_2})}{\beta_{\mathrm{HCO^+}} n(\mathrm{e}) (k_{\mathrm{H_3^+}} n(\mathrm{CO}) + n(\mathrm{e}) \beta_{\mathrm{H_3^+}} )}.
\end{equation}
Here $\zeta$ is the ionization rate, $n(\mathrm{e})$ is the electron number density, $k_{\mathrm{H_3^+}}$ is the destruction rate for \hhhp\ reacting with CO, 
$\beta_{\mathrm{HCO^+}}$ and $\beta_{\mathrm{H_3^+}}$ are the electron recombination rates for \hcop and \hhhp\ respectively.
Equation~\ref{eq:final} includes four unknowns: the ionization rate, the electron number density, the CO number density, and the \hh\ number density.

We calculate the \hcop\ and CO column densities based on the observations of \hthcop\ J = 1-0 and \cooo\ J = 2-1, assuming the lines are optically thin. 
Since \cooo\ was observed at higher spatial resolution than \hthcop, we first convolve the \cooo\ fits cube with the \hthcop\ beam using the Python package \texttt{imgcube}\footnote{\url{https://github.com/richteague/imgcube}}, then regrid the pixels in the \cooo\ data to match the \hthcop\ pixels using a linear interpolation.
Following the method of \citet{Goldsmith99}, we calculate the column density from:
\begin{equation}
    N_u = \frac{8\pi k \nu_{0}^{2} W}{h c^3 A_{ul}},
\end{equation}
where $N_u$ is the upper state column density, $\nu_0$ is the rest frequency of the line, $W$ is the integrated line intensity in K km s$^\mathrm{-1}$, and $A_{ul}$ is the Einstein A coefficient of the transition.  
The upper state column density is only calculated in pixels where the peak flux is at least three times the rms noise level.

\begin{figure*}
    \centering
    \includegraphics[width=2\columnwidth]{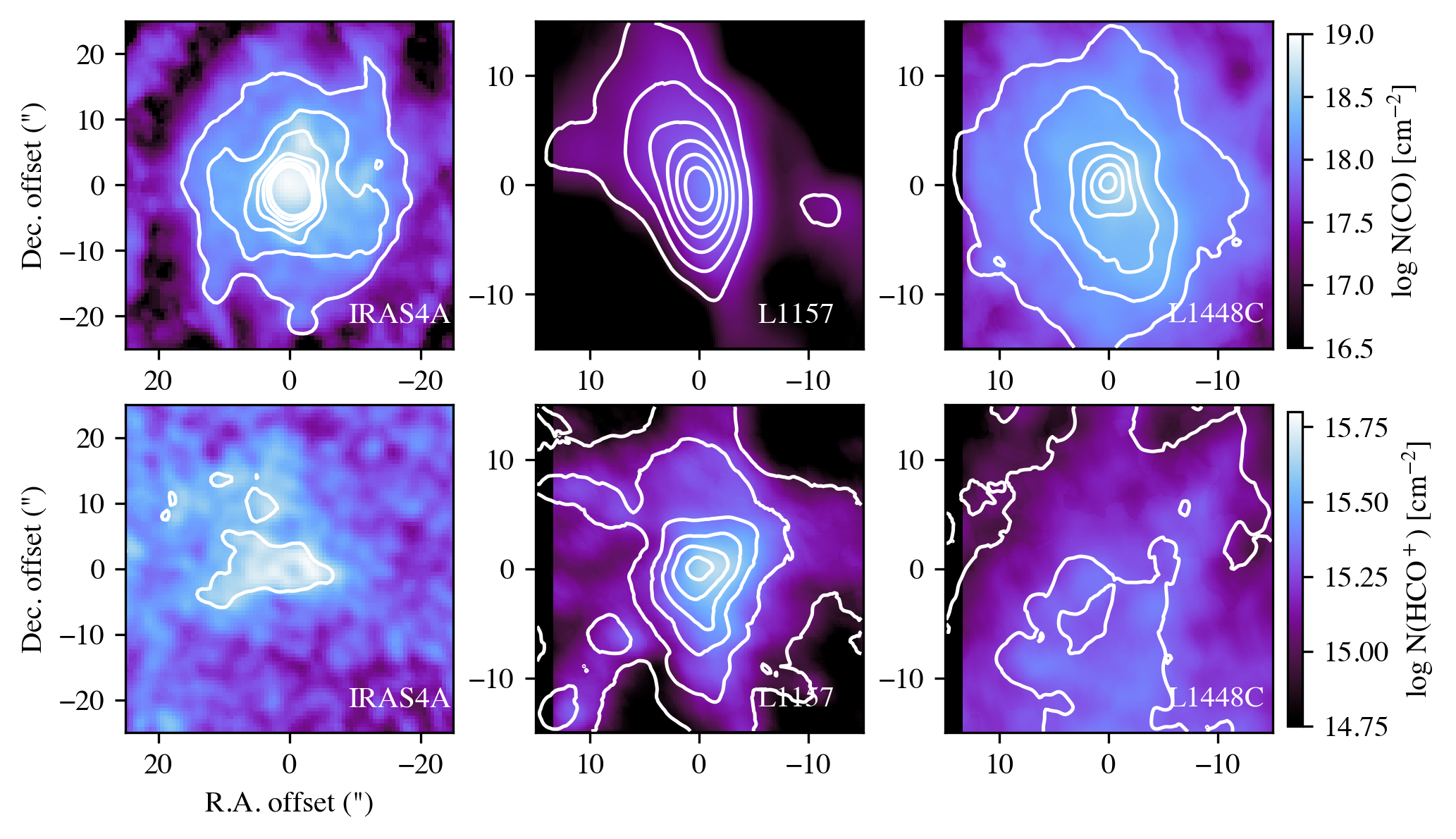}
    \caption{Maps of the derived column densities for CO (top) and \hcop\ (bottom) based on our observations of \cooo\ and \hthcop\ respectively. Contours indicate the isotopologue emission from 3 to 21 times the RMS in 5 equally spaced steps.}
    \label{fig:Nmap}
\end{figure*}

To convert the upper state column density to total column density, $N$, we assume local thermodynamic equilibrium:
\begin{equation}
    N = N_u \frac{Q}{g_u} \mathrm{e}^{E_u/T_{ex}},
\end{equation}
where $Q$ is the rotational partition function for a given excitation temperature $T_{ex}$, $g_u$ is upper state statistical weight, and $E_u$ is the upper state energy.
We used the excitation temperatures from the source specific \cooo\ models of \citet[][their Table~3]{Anderl16}. The assumed temperatures are 28~K for IRAS4A, 27~K for L1448-C, and 26~K for L1157, appropriate for a radius of 400 au in a sphereical envelope. This corresponds roughly to the spatial resolution of our observations. 
The envelope gradually cools at larger radii. For example, the model temperature profile for IRAS4A is 28~K at 560 au and 10~K at 5000 au \citep{Notsu21}. Thus our assumed temperature should be considered an upper limit at large offsets, resulting in a slight overestimate of the derived column densities. Given their different partition functions, this overestimate is larger for \hcop\ than for CO. Ultimately, decreasing the assumed temperature changes the derived ionization rates by roughly a factor of three to four.
Finally, to convert from isotopologue abundance to total abundance we assume abundance ratios of 69 for $\mathrm{^{12}C/^{13}C}$ and 557 for $\mathrm{^{16}O/^{18}O}$, based on measurements of the local interstellar medium \citep{Wilson99}. Maps of the derived CO and \hcop\ column densities are shown in Figure~\ref{fig:Nmap}. As a quick check we compare the column densities we derive for the L1157 envelope to those for the same species in the L1157 outflow \citep{Podio14} and find columns several orders of magnitude greater in the envelope, as expected for denser gas.

\begin{figure*}
    \centering
    \includegraphics[width=2\columnwidth]{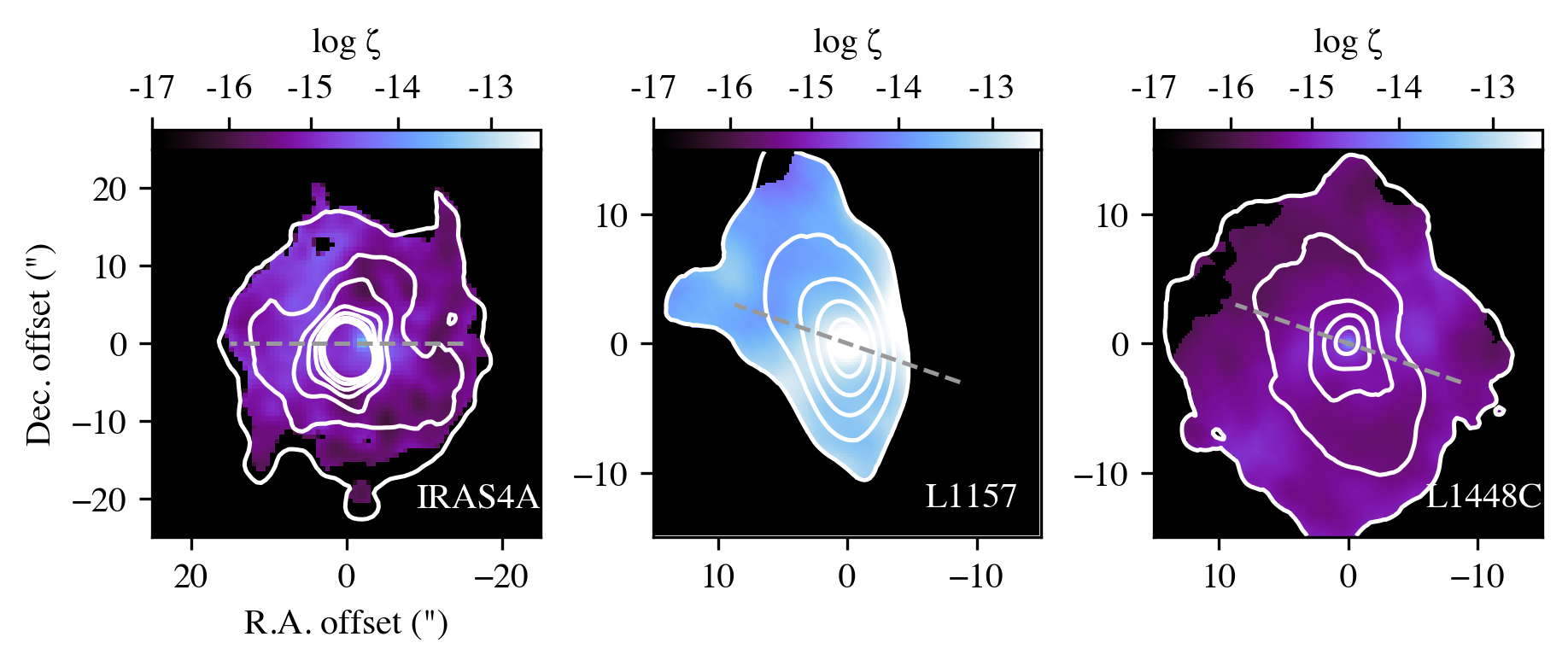}
    \caption{Cosmic ray ionization rate derived from the abundance ratio of \hcop\ and CO assuming an electron abundance equal to the \hcop\ abundance \citep{Caselli02}. The ionization rate is calculated only in regions where both tracers are detected at $>3\sigma$. Contours indicate the \cooo\ 2-1 emission, from 3 to 21 times the RMS in 5 equally spaced steps. The \cooo\ for IRAS4A is shown after correcting for absorption. In each panel, the grey dashed line is perpendicular to the small scale outflow.}
    \label{ion}
\end{figure*}

To solve for $\zeta$ in Equation~\ref{eq:hcop}, assumptions must be made regarding the electron, CO, and \hh\ number densities. For the \hh\ density, we assume a spherical cloud with a power-law density profile, using the source specific parameters fit by \citet{Kristensen12}. The projected distance from source center is used to obtain the density at a given position.
We then set the CO number density using the best fit CO/\hh\ ratio for each source from the models of \citet{Anderl16}.
We make the conservative assumption that the number density of electrons must be at least equal to the number density of \hcop \citep[e.g.,][]{vantHoff18}.
The derived number densities, column densities, and ionization rate for a representative offset of 2.5$\arcsec$ perpendicular to the outflow are given in Table~\ref{tab:derivedprops}.
The resulting ionization rate maps are shown in Figure~\ref{ion}.

\begin{deluxetable*}{lcccccc}[!htbp]
\tablecaption{Assumed and Derived properties offset $2.5\arcsec$ from source center and perpendicular to the outflow axis. Uncertainties are the standard deviation of the negative and positive offset.}
\label{tab:derivedprops}
\tablehead{
\colhead{Source} & \colhead{n(\hh)$^{(a)}$} & \colhead{X(CO)$^{(b)}$}& 
\colhead{T$_\mathrm{ex}$$^{(b)}$} &
\colhead{N(CO)} & \colhead{N(HCO$^\mathrm{+}$)} & \colhead{$\zeta$} \\  
 & \colhead{(cm$^{-3}$)} &  & \colhead{(K)} &  \colhead{(cm$^{-2}$)} & \colhead{(cm$^{-2})$} & \colhead{(s$^{-1}$)} 
}
\startdata 
IRAS4A & $1.53 \ee{7}$ & $1.114\ee{-5}$ & 28  & $6.4\pm1.6\ee{18}$  & $5.5\pm0.3\ee{15}$  & $1.21\pm0.03\ee{-15}$  \\ 
L1448-C & $8.25\ee{6}$ & $1.114\ee{-5}$ & 27 & $3.13\pm0.09\ee{18}$ & $1.9\pm0.2\ee{15}$ & $6.1\pm0.7\ee{-16}$ \\ 
L1157 & $3.26 \ee{6}$ & $3.342\ee{-5}$ & 26 & $9.4\pm0.3\ee{17}$  & $3.65\pm0.03\ee{15}$ & $1.1\pm0.2\ee{-13}$  \\ 
\enddata 
\tablecomments{$^{(a)}$ from \citet{Kristensen12}. $^{(b)}$ from \citet{Anderl16}}
\end{deluxetable*}

\section{Results and Discussion}
Using observations of \hthcop\ and \cooo\ in three Class 0 protostellar systems, we have constrained the cosmic ray ionization rate for an assumed electron abundance (Figure~\ref{ion}). In all three systems the maximum ionization rate exceeds both the value of 1.3\ee{-17} s$^{-1}$ from spacecraft and the updated diffuse ISM value of 6\ee{-17} s$^{-1}$ by several orders of magnitude \citep{Webber98,Obolentseva24}. 

For all three protostellar sources, the derived ionization rate peaks at the same position as the \cooo\ flux, where \cooo\ is most likely to be optically thick. As a quick check, we calculate the optical depth of the \cooo\ 2-1 emission in the central pixel of the IRAS4A map using the assumed gas temperature and \hh\ number density from Section~\ref{sec:ana} using \texttt{RADEX} \citep{vanderTak07}. We are able to reproduce the observed peak flux with an optically thin line ($\tau = 0.07$). However, our assumed source properties are highly uncertain near the protostar due to a lack of sufficiently high spatial resolution observations.
If the emission is in fact optically thick, our optically thin assumption underestimates the total CO column density. 
By extension the derived ionization rate at source center is likely an overestimate of the true value. Additionally, the power-law profile used to describe the \hh\ density is uncertain in the innermost regions of the protostar, as it is mainly constrained by dust continuum observations at spatial scales larger than 10\as\ \citep{Kristensen12}.

Beyond the central pixels, the derived envelope ionization rate is still enhanced, of order $\zeta = \eten{-16}-\eten{-15}$ s$^{-1}$ for IRAS4A and L1448-C, and $\zeta = \eten{-14}-\eten{-13}$ s$^{-1}$ for L1157. These results are consistent with the enhanced ionization rate toward OMC-2 FIR4 based on analysis of the \hcop\ and \nnhp\ emission in the infrared \citep{Ceccarelli14,Favre17}.
The derived ionization rate for L1157 is noticeably higher than for the other sources in our sample. Ultimately, this is the result of a higher \hthcop\ to \cooo\ flux ratio for L1157 compared to the other sources. Given the small sample size, it is unclear if this is due to any particular source property. A larger survey is needed to determine the true spread in ionization rates of protostellar envelopes. 

\citet{Podio14} constrained in the ionization rate at the bow shock L1157-B1 in the L1157 outflow to be $\zeta=3\ee{-16}$ s$^{-1}$ based on \hcop\ and \nnhp\ observations from the IRAM 30m. This rate was also found to be consistent with nitrogen fractionation in HCN isotopologues in the B1 knot \citep{Benedettini21}. This is significantly lower than the ionization rate for the L1157 envelope derived in the current work. There are several potential explanations for this apparent mismatch. The current work employs an analytical approach to derive the ionization rate, while \citet{Podio14} employ a chemical model to reproduce the observed line fluxes, where the maximum ionization rate considered is $\zeta=3\ee{-16}$ s$^{-1}$. \citet{Podio14} also conclude that the \hcop\ and \nnhp\ abundances in L1157-B1 are not changed by the shock but rather reflect the chemistry of the pre-shock gas \citep{Codella13}. In this interpretation, these ions have not been influenced by any local ionization enhancement within the shock.

At the same time, it is possible that the current work over-predicts the ionization rate in the L1157 envelope. In particular, if our assumption of optically thin emission is incorrect and the \cooo\ emission is more optically thick than the \hthcop\ emission, Equation~3 would overestimate the ionization rate. However, it should be noted that L1157 has the lowest derived CO column density of our three sources and is therefore the least likely to be optically thick. A detailed study of the L1157 system is required to put the ion emission from envelope, outflow, and shock into their proper context.

Our determination of the ionization rate in these three Class 0 protostellar envelopes is based on an analytical calculation for the assumed \hh\ densities. In prestellar cores, where CO is largely frozen out as ice, \hhdp\ is used to trace the ionization rate \citep{Sabatini23,Redaelli25}. In these objects the \hcop/CO ratio is not a robust tracer of ionization \citep{Redaelli24}.
In protostellar envelopes \hhdp\ emission arises from colder regions of the envelope compared to \hcop\ and CO, as shown for IRAS4A \citep{Koumpia17}, and could potentially be used to constrain ionization in this region.
It is important for observers to consider the physical conditions in their targets when choosing an ionization tracer.

More precise constraints on the envelope ionization rate require detailed chemical modeling, similar to previous analysis of Class II sources and hot cores \citep[e.g.,][]{Cleeves15,Barger20}. Chemical modeling is also required to determine the source of the enhanced ionization. UV, X-rays, and cosmic rays can all ionize molecular gas. 
The X-ray luminosities of Class 0 protostars are difficult to constrain observationally due to absorption by the surrounding envelope \citep{Gudel09}. However, chemical modeling demonstrates that X-rays can increase the production of \hcop\ within a few hundred au of the protostar \citep{Notsu21}.
The UV luminosities of embedded protostars are thought to be higher than for Class II sources \citep{Visser12}. 
Shocks can also produce both UV photons and cosmic rays \citep{Karska18}. The UV radiation could contribute to enhanced ionization within the outflow cavity. However, protostellar envelopes are optically thick to UV radiation. The UV field will not contribute to any ionization driven chemistry deep in the envelope or embedded disk. Cosmic rays, on the other hand, can penetrate much deeper, potentially providing a source of chemistry-driving ionization to the youngest circumstellar disks.

\section{Summary}
We used new observations of \hthcop\ J = 1-0 towards three protostars, combined with archival observations of \cooo\ J = 2-1 from the CALYPSO program to measure the ionization rate in the protostellar envelope. For all three sources we find ionization rates well above the average diffuse ISM value of 6\ee{-17} s$^{-1}$, in the range \eten{-16}-\eten{-13} s$^{-1}$. 
Detailed chemical modeling is required to determine if the ionization is due to X-rays, UV photons, or cosmic rays. 
These results are consistent with previous findings of a high ionization rate in two additional protostellar envelopes. Observations of more sources are needed to determine if enhanced ionization is a common occurrence at the protostellar stage. Additionally, more detailed modeling of the chemistry in protostellar envelopes is needed to provide more accurate constraints on the ionization rate, as well as the relative contribution of cosmic rays and high energy photons. If enhanced ionization is found to be common, it will favor a scenario in which the bulk of chemical processing occurs through ionization-driven chemistry during the Class 0 phase. 

\begin{acknowledgements}
The authors thank the IRAM staff at the NOEMA observatory and IRAM 30 m telescope for their support in the observations and data calibration.
This work is based on observations carried out under project numbers U052, S16AD and W17AC with the IRAM NOEMA Interferometer and project number 010-22 with the IRAM 30 m telescope. IRAM is supported by INSU/CNRS (France), MPG (Germany) and IGN (Spain).
\end{acknowledgements}

\facility{NOEMA, IRAM 30m telescope}
\software{GILDAS, matplotlib \citep{matplotlib}, numpy \citep{numpy}, emcee \citep{emcee}}

\bibliography{ionletter}{}
\bibliographystyle{aasjournalv7}

\end{document}